\documentclass[aps,prd,showpacs,superscriptaddress,twocolumn]{revtex4} % UTOBS

\usepackage{graphicx}
\usepackage{natbib}
\usepackage{slashed}

\begin{document}
%\pagewiselinenumbers  % LINENO CTOBS

\preprint{CDF/DOC/TOP/CDFR/8480} %Comment this out before shipping out
\preprint{Phys. Rev. D draft 2.1}
%\vspace*{-1.5cm}
%\includegraphics[width=0.15\textwidth]{Pictures/cdf_ii.eps}
%\vspace{1.5cm}

\title{Measurement of the Helicity Fractions of W Bosons from Top Quark 
       Decays using Fully Reconstructed \boldmath${t\bar{t}}$ Events with CDF II}

\vspace{0.5in}

%This gets replaced by authors and institutions
%\author{The CDF Collaboration} % This will get replaced with the CDF author list. See the GP webpage to get it.

\affiliation{Institute of Physics, Academia Sinica, Taipei, Taiwan 11529, Republic of China} 
\affiliation{Argonne National Laboratory, Argonne, Illinois 60439} 
\affiliation{Institut de Fisica d'Altes Energies, Universitat Autonoma de Barcelona, E-08193, Bellaterra (Barcelona), Spain} 
\affiliation{Baylor University, Waco, Texas  76798} 
\affiliation{Istituto Nazionale di Fisica Nucleare, University of Bologna, I-40127 Bologna, Italy} 
\affiliation{Brandeis University, Waltham, Massachusetts 02254} 
\affiliation{University of California, Davis, Davis, California  95616} 
\affiliation{University of California, Los Angeles, Los Angeles, California  90024} 
\affiliation{University of California, San Diego, La Jolla, California  92093} 
\affiliation{University of California, Santa Barbara, Santa Barbara, California 93106} 
\affiliation{Instituto de Fisica de Cantabria, CSIC-University of Cantabria, 39005 Santander, Spain} 
\affiliation{Carnegie Mellon University, Pittsburgh, PA  15213} 
\affiliation{Enrico Fermi Institute, University of Chicago, Chicago, Illinois 60637} 
\affiliation{Comenius University, 842 48 Bratislava, Slovakia; Institute of Experimental Physics, 040 01 Kosice, Slovakia} 
\affiliation{Joint Institute for Nuclear Research, RU-141980 Dubna, Russia} 
\affiliation{Duke University, Durham, North Carolina  27708} 
\affiliation{Fermi National Accelerator Laboratory, Batavia, Illinois 60510} 
\affiliation{University of Florida, Gainesville, Florida  32611} 
\affiliation{Laboratori Nazionali di Frascati, Istituto Nazionale di Fisica Nucleare, I-00044 Frascati, Italy} 
\affiliation{University of Geneva, CH-1211 Geneva 4, Switzerland} 
\affiliation{Glasgow University, Glasgow G12 8QQ, United Kingdom} 
\affiliation{Harvard University, Cambridge, Massachusetts 02138} 
\affiliation{Division of High Energy Physics, Department of Physics, University of Helsinki and Helsinki Institute of Physics, FIN-00014, Helsinki, Finland} 
\affiliation{University of Illinois, Urbana, Illinois 61801} 
\affiliation{The Johns Hopkins University, Baltimore, Maryland 21218} 
\affiliation{Institut f\"{u}r Experimentelle Kernphysik, Universit\"{a}t Karlsruhe, 76128 Karlsruhe, Germany} 
\affiliation{High Energy Accelerator Research Organization (KEK), Tsukuba, Ibaraki 305, Japan} 
\affiliation{Center for High Energy Physics: Kyungpook National University, Taegu 702-701, Korea; Seoul National University, Seoul 151-742, Korea; and SungKyunKwan University, Suwon 440-746, Korea} 
\affiliation{Ernest Orlando Lawrence Berkeley National Laboratory, Berkeley, California 94720} 
\affiliation{University of Liverpool, Liverpool L69 7ZE, United Kingdom} 
\affiliation{University College London, London WC1E 6BT, United Kingdom} 
\affiliation{Centro de Investigaciones Energeticas Medioambientales y Tecnologicas, E-28040 Madrid, Spain} 
\affiliation{Massachusetts Institute of Technology, Cambridge, Massachusetts  02139} 
\affiliation{Institute of Particle Physics: McGill University, Montr\'{e}al, Canada H3A~2T8; and University of Toronto, Toronto, Canada M5S~1A7} 
\affiliation{University of Michigan, Ann Arbor, Michigan 48109} 
\affiliation{Michigan State University, East Lansing, Michigan  48824} 
\affiliation{Institution for Theoretical and Experimental Physics, ITEP, Moscow 117259, Russia} 
\affiliation{University of New Mexico, Albuquerque, New Mexico 87131} 
\affiliation{Northwestern University, Evanston, Illinois  60208} 
\affiliation{The Ohio State University, Columbus, Ohio  43210} 
\affiliation{Okayama University, Okayama 700-8530, Japan} 
\affiliation{Osaka City University, Osaka 588, Japan} 
\affiliation{University of Oxford, Oxford OX1 3RH, United Kingdom} 
\affiliation{University of Padova, Istituto Nazionale di Fisica Nucleare, Sezione di Padova-Trento, I-35131 Padova, Italy} 
\affiliation{LPNHE, Universite Pierre et Marie Curie/IN2P3-CNRS, UMR7585, Paris, F-75252 France} 
\affiliation{University of Pennsylvania, Philadelphia, Pennsylvania 19104} 
\affiliation{Istituto Nazionale di Fisica Nucleare Pisa, Universities of Pisa, Siena and Scuola Normale Superiore, I-56127 Pisa, Italy} 
\affiliation{University of Pittsburgh, Pittsburgh, Pennsylvania 15260} 
\affiliation{Purdue University, West Lafayette, Indiana 47907} 
\affiliation{University of Rochester, Rochester, New York 14627} 
\affiliation{The Rockefeller University, New York, New York 10021} 
\affiliation{Istituto Nazionale di Fisica Nucleare, Sezione di Roma 1, University of Rome ``La Sapienza," I-00185 Roma, Italy} 
\affiliation{Rutgers University, Piscataway, New Jersey 08855} 
\affiliation{Texas A\&M University, College Station, Texas 77843} 
\affiliation{Istituto Nazionale di Fisica Nucleare, University of Trieste/\ Udine, Italy} 
\affiliation{University of Tsukuba, Tsukuba, Ibaraki 305, Japan} 
\affiliation{Tufts University, Medford, Massachusetts 02155} 
\affiliation{Waseda University, Tokyo 169, Japan} 
\affiliation{Wayne State University, Detroit, Michigan  48201} 
\affiliation{University of Wisconsin, Madison, Wisconsin 53706} 
\affiliation{Yale University, New Haven, Connecticut 06520} 
\author{A.~Abulencia}
\affiliation{University of Illinois, Urbana, Illinois 61801}
\author{J.~Adelman}
\affiliation{Enrico Fermi Institute, University of Chicago, Chicago, Illinois 60637}
\author{T.~Affolder}
\affiliation{University of California, Santa Barbara, Santa Barbara, California 93106}
\author{T.~Akimoto}
\affiliation{University of Tsukuba, Tsukuba, Ibaraki 305, Japan}
\author{M.G.~Albrow}
\affiliation{Fermi National Accelerator Laboratory, Batavia, Illinois 60510}
\author{D.~Ambrose}
\affiliation{Fermi National Accelerator Laboratory, Batavia, Illinois 60510}
\author{S.~Amerio}
\affiliation{University of Padova, Istituto Nazionale di Fisica Nucleare, Sezione di Padova-Trento, I-35131 Padova, Italy}
\author{D.~Amidei}
\affiliation{University of Michigan, Ann Arbor, Michigan 48109}
\author{A.~Anastassov}
\affiliation{Rutgers University, Piscataway, New Jersey 08855}
\author{K.~Anikeev}
\affiliation{Fermi National Accelerator Laboratory, Batavia, Illinois 60510}
% ...
\collaboration{CDF Collaboration\footnote{With visitors from $^a$University of Athens, 
$^b$University of Bristol, 
$^c$University Libre de Bruxelles, 
$^d$Cornell University, 
$^e$University of Cyprus, 
$^f$University of Dublin, 
$^g$University of Edinburgh, 
$^h$University of Heidelberg, 
$^i$Universidad Iberoamericana, 
$^j$University of Manchester, 
$^k$Nagasaki Institute of Applied Science, 
$^l$University de Oviedo, 
$^m$University of London, Queen Mary and Westfield College, 
$^n$Texas Tech University, 
$^o$IFIC(CSIC-Universitat de Valencia), 
}}
\noaffiliation

\date{\today}

\begin{abstract}
We present a measurement of the fractions $F_0$ and $F_+$ of longitudinally polarized
and right-handed $W$ bosons in top quark decays
using data collected with the CDF II detector.
The data set used in the analysis corresponds to an integrated luminosity of 
approximately $318\;\mbox{pb}^{-1}$. 
We select $t\bar{t}$ candidate events with one lepton, at least four jets, and missing
transverse energy.
Our helicity measurement uses the decay angle $\theta^*$, which is defined as the
angle between the momentum of the charged lepton 
in the $W$ boson rest frame and the $W$ momentum in the top quark rest frame.
The $\cos\theta^*$ distribution in the data is determined by full
kinematic reconstruction of the $t\bar{t}$ candidates.
We find $F_0=0.85^{+0.15}_{-0.22}(\mathrm{stat})\pm 0.06\,(\mathrm{syst})$ and 
$F_+=0.05^{+0.11}_{-0.05}(\mathrm{stat})\pm 0.03\,(\mathrm{syst})$,
which is consistent with the standard model prediction.
We set an upper limit on the fraction of right-handed $W$ bosons
of $F_+ < 0.26$ at the 95\% confidence level.
\end{abstract}

% 12.15.-y = Electroweak interactions
% 13.88.+e = Polarization in interactions and scattering
% 14.65.Ha = Top quarks
% 14.70.Fm = W bosons
%\keywords{}
%
% Use noaffiliation if just using the CDF collaboration as author.
\noaffiliation

\maketitle


\section{Introduction}
\label{sec:introduction}
In 1995 the top quark was discovered at the Tevatron proton-antiproton 
collider at Fermilab by the CDF and D\O\ 
collaborations~\cite{Abe:1995hr,Abachi:1995iq}.
It is the most massive known elementary particle and its mass
is currently measured with a precision of about 
1.3\%~\cite{Group:2006qt,Abulencia:2005ak}.
However, the measurements of other top quark properties are still statistically limited, so 
the question remains whether the standard model successfully predicts these properties.
This paper addresses one interesting aspect
of top quark decay, the helicity of the $W$ boson produced in the decay $t\rightarrow W^+\, b$.
 
At the Tevatron collider, with a center-of-mass energy $\sqrt{s}=1.96\;\mbox{TeV}$,
most top quarks are pair-produced via the strong 
interaction. In the standard model the top quark decays predominantly
into a $W$ boson and a $b$ quark, with a branching ratio close
to 100\%.
The $V-A$ structure of the weak interaction of the standard model 
predicts that the $W^+$ bosons from the top quark decay 
$t\rightarrow W^+\, b$
are dominantly either longitudinally polarized or left-handed, while right-handed
$W$ bosons are heavily suppressed and are forbidden in the limit
of massless $b$ quarks.

As a consequence of the Goldstone boson equivalence 
theorem~\cite{Cornwall:1974km,Lee:1977eg},
the decay amplitude to longitudinal $W$ bosons is proportional to the
Yukawa coupling of the top quark; therefore, the decay rate scales with $m_t^3$,
where $m_t$ is the top quark mass. 
The longitudinal decay mode of the $W$ boson is thereby linked
to the spontaneous breaking of the electroweak gauge symmetry. 
The decay rate to transverse
$W$ bosons is governed by the gauge coupling and increases only
linearly with $m_t$~\cite{Kuhn:1996ug}.
The fraction of longitudinally polarized $W$ bosons is defined by
\begin{equation}
\label{eq:f0}
F_0 = \frac{\Gamma(t\rightarrow W_0^+b)}{\Gamma(t\rightarrow W_{\rm L}^+b)+
  \Gamma(t\rightarrow W_{\rm 0}^+b)+\Gamma(t\rightarrow W_{\rm R}^+b)}\;,
\end{equation}
where $W_0^+$ stands for a longitudinally polarized $W^+$ boson,
$W_{\rm L}^+$ for a left-handed $W^+$ boson, and $W_{\rm R}^+$ for a right-handed
$W^+$ boson. The corresponding definitions for the $W^-$ boson are 
implied.
In leading-order perturbation theory $F_0$ is predicted to be
$F_0=\frac{m_t^2}{2m_W^2+m_t^2}$~\cite{Kane:1991bg}, where
$m_W$ is the mass of the $W$ boson.
Using $m_W=80.43\;\mbox{GeV}/c^2$~\cite{Eidelman:2004wy} and
$m_t=(172.5\pm2.3)\;\mbox{GeV}/c^2$~\cite{Group:2006qt},
gives $F_0=0.697\pm0.007$, where the given uncertainty is
only due to the uncertainty in the top quark mass. 
% +1 sigma: 0.703
% -1 sigma: 0.691
Next-to-leading-order corrections to the total decay width
and the partial decay width into longitudinal $W$ bosons
amount to about -10\%~\cite{jezabekKuehn1989,jezabekKuehn1988,czarnecki1999,chetyrkin1999,denner1991,migneron1991,Fischer:1998gs,Fischer:2000kx,Fischer:2001gp,Do:2002ky}.
However, the fraction of longitudinal $W$ bosons is only 
negligibly changed.

A significant deviation from the predicted value for $F_0$ or a 
nonzero value for the
right-handed fraction $F_+$ could indicate new physics.
Left-right symmetric models~\cite{Peccei:1990uv}, 
for example, 
lead to a significant right-handed fraction of $W$
bosons in top quark decays. Such a right-handed component
($V+A$ coupling) would lead to a smaller left-handed
fraction, while the longitudinal fraction $F_0$ would change
insignificantly.
Since the decay rate to longitudinal $W$ bosons depends on the
Yukawa coupling of the top quarks, the measurement of $F_0$ is
sensitive to the mechanism of electroweak symmetry breaking.
Alternative models for electroweak symmetry breaking, such as
topcolor-assisted technicolor models, can lead to an altered $F_0$ 
fraction~\cite{Wang:2005ra,Chen:2005vr}. 

The $W$ boson polarization manifests itself in the decay $W\rightarrow \ell\nu$ in the angle $\theta^*$, which is defined as the angle between the momentum of the charged lepton 
in the $W$ rest frame and the $W$ momentum in the top quark rest frame.
For a longitudinal fraction $F_0$, a right-handed fraction $F_+$,
and a left-handed fraction $F_-=1-F_+ - F_0$, the $\cos\theta^*$ distribution is given 
by~\cite{Kane:1991bg}:
\begin{eqnarray}
\frac{dN}{d\cos\theta^*} &=& (1-F_+ - F_0)\cdot\frac{3}{8}(1-\cos\theta^*)^2 \nonumber \\
&+& (F_0)\cdot\frac{3}{4}(1-\cos^2\theta^*) \nonumber \\
&+& (F_+)\cdot\frac{3}{8}(1+\cos\theta^*)^2 .
\label{cos_general}
\end{eqnarray}

In this analysis, the $W$ helicity fractions are measured in a selected sample rich in $t\bar{t}$ events where one lepton, at least four jets, and missing transverse energy are required~\cite{Abulencia:2006in}.
In order to calculate $\theta^*$, all kinematic quantities describing the $t\bar{t}$ decays
have to be determined.

Previous CDF measurements of the $W$ helicity fractions in top quark
decays used either the square of the invariant mass of the charged lepton 
and the $b$ quark jet~\cite{Acosta:2004mb, Abulencia:2005xf, Abulencia:2006iy} or the lepton $p_\mathrm{T}$ 
distribution~\cite{Affolder:1999mp, Abulencia:2005xf} as a discriminant.
The D\O \ collaboration used a matrix-element method to extract a value for $F_0$~\cite{Abazov:2004ym};
in a second analysis the reconstructed distribution of $\cos\theta^*$~\cite{Abazov:2005fk} 
was utilized to measure $F_+$.
The previous measurement by CDF was $F_{0} = 0.74^{+0.22}_{-0.34}$~\cite{Abulencia:2005xf}, while D\O\ measured $F_{0} = 0.56\pm0.31$~\cite{Abazov:2004ym}.
The CDF collaboration also measured the current best upper limit of $F_{+} < 0.09$ at the 95\% confidence level~\cite{Abulencia:2006iy}.

The organization of this paper is as follows. Section II describes the detector system relevant to this analysis. 
Section III illustrates the event selection of the $t\bar{t}$ candidates. The signal simulation and background estimation are given in Section IV. In Section V we describe our method to fully reconstruct $t\bar{t}$ pairs. The extraction of the helicity fractions is presented in Section VI. Section VII discusses the systematic uncertainties. Finally, the results and conclusions are given in Section VIII.

\section{The CDF II Detector}
% ttbar cross section: Event Detection and Reconstruction

A detailed description of the Collider Detector at
Fermilab (CDF) can be found elsewhere~\cite{Acosta:2004yw}.
A coordinate system with the $z$ axis along the proton beam,
azimuthal angle $\phi$, and polar angle $\theta$ is used. 
The azimuthal angle is defined with respect to the outgoing radial direction and the polar angle
is defined with respect to the proton beam direction.
The transverse energy of a particle is defined as $E_{\rm T}=E\sin\theta$.
Throughout this paper we use pseudorapidity defined as $\eta = -\ln(\tan\frac{\theta}{2})$. 
The primary detector components relevant to this analysis
are those which measure the energies and directions of jets, electrons, and muons and are briefly described below. 

An open-cell drift chamber, the central outer tracker (COT)~\cite{Affolder:2003ep}, 
and a silicon tracking system are used to measure the 
momenta of charged particles.
The CDF II silicon tracker consists of three subdetectors:
a layer of single-sided silicon microstrip
detectors~\cite{Hill:2004qb} glued on the beam pipe, a five layer double-sided
silicon microstrip detector (SVX~II)~\cite{Sill:2000zz}, and 
intermediate silicon layers~\cite{Affolder:2000tj} located at radii
between 19 and $29\,\mathrm{cm}$ which provide linking
between track segments in the COT and the SVX~II.
In the analysis presented in this article, the silicon 
tracker is used to identify jets
originating from $b$ quarks by reconstructing secondary vertices.
The tracking detectors are located within a 1.4 T solenoid.
Electromagnetic and hadronic sampling calorimeters~\cite{Balka:1987ty, Bertolucci:1987zn,Albrow:2001jw}, which have an angular coverage of $|\eta|<3.6$, surround the tracking system and measure the energy flow of interacting particles.
They are segmented into projective towers, each one covering a small range in pseudorapidity and azimuth. 
For electron identification the electromagnetic calorimeters are used, while jets are identified 
through the energy they deposit in the electromagnetic and hadronic calorimeter towers.
The muon system~\cite{Ascoli:1987av} is located outside of the calorimeters and provides muon detection in the range $|\eta|<1.5$.
Muons penetrating the five absorption lengths of the calorimeters are detected in planes of multi-wire drift chambers.
Since the collision rate exceeds the tape writing speed by five orders of magnitude, CDF has a three-level trigger system which reduces the event rate from 
1.7~MHz to 60~Hz for permanent storage.
The first two levels of trigger are implemented by special-purpose hardware, whereas the third one is implemented by software running on a computer farm.

\section{Selection of \boldmath${t\bar{t}}$ Candidate Events}
\label{sec:eventSel}

In the decay channel considered in this analysis, one top quark decays semileptonically and the second top quark
decays hadronically, leading to a signature of one charged lepton, missing transverse energy resulting from the undetected neutrino, and at 
least four jets.
Candidate events are selected with high-$p_{\rm T}$ lepton triggers. The electron trigger requires a COT track matched to an energy cluster
in the central electromagnetic calorimeter with $E_{\rm T}>$ 18~GeV. The muon trigger requires a COT track with $p_{\rm T}>$ 18~GeV$/c$ matched to a track 
segment in the muon chambers.
After offline reconstruction, we require exactly one isolated electron candidate with $E_{\rm T}>20\,\mathrm{GeV}$ and $|\eta|<1.1$
or exactly one isolated muon candidate with $p_\mathrm{T}>20\,\mathrm{GeV}/c$ and $|\eta|<1.0$.
An electron or muon candidate is considered isolated if the $E_{\rm T}$ not assigned to the lepton in a cone of $R \equiv \sqrt{(\Delta \eta)^2+ (\Delta \phi)^2}=0.4$  
centered around the lepton is less than 10\% of the lepton $E_{\rm T}$ or $p_{\rm T}$, respectively.
Jets are reconstructed by summing calorimeter energy in a cone of radius $R = 0.4$. 
The energy of the jets is corrected~\cite{Bhatti:2005ai} for the $\eta$ dependence of the calorimeter response, the time dependence of the calorimeter response, and the extra deposition of energy due to multiple interactions.
Candidate jets must have corrected $E_{\rm T} > 15 \;\mbox{GeV}$ and detector $|\eta| < 2.0$. Detector $\eta$ is defined as the pseudorapidity
of the jet calculated with respect to the center of the detector.
Events with at least four jets are accepted. 
At least one of the jets must be tagged as a $b$-jet by requiring a displaced secondary vertex within the jet~\cite{Acosta:2004hw}.
The missing $E_{\rm T}$ ($\not\!\! \vec{E}_{\rm T}$) is defined by
\begin{eqnarray}
\not\!\! \vec{E}_{\rm T} & = & - \sum_{i} E_{\rm T}^i \hat{n}_i, \\
~i& = & \rm calorimeter~tower~number~with~|\eta| < 3.6, \nonumber 
\end{eqnarray}
where $\hat{n}_i$ is a unit vector perpendicular to the beam axis and pointing at the i$^{th}$ calorimeter tower. We also define $\not\!\! E_{\rm T} = |\not\!\! \vec{E}_{\rm T}|$.
Because this calculation is based on calorimeter towers, $\not\!\! E_{\rm T}$ has to be adjusted for the effect of the jet corrections for all jets with 
$E_{\rm T}>$ 8~GeV and detector $\left | \eta \right |<$ 2.5.
In events with muons, the transverse momentum of the muon is added to the sum, and a correction is applied to remove the average ionization 
energy released by the muon in traversing the calorimeter. 
We require the corrected $\not\!\! E_{\rm T}$ to be greater than 20~GeV.

Additional requirements reduce the contamination from background.
Electron events are rejected if the electron stems from a conversion of a photon.
Cosmic ray muon events are also excluded.
To remove $Z$ boson events, we reject events in which the charged lepton can be paired with any more loosely defined jet or lepton to form an invariant
mass consistent with the $Z$ peak, defined as the range 76~GeV/$c^2$ to 106~GeV/$c^2$.
After these selection requirements we find 82 $t\bar{t}$ candidates
in the selected sample corresponding to an integrated luminosity of $318\;\mbox{pb}^{-1}$.

\section{Signal Simulation and Background Estimation}
\label{sec:background}

In order to determine the resolution of the kinematic quantities of the reconstructed $t\bar{t}$ pair, as well as
to determine certain background rates, we utilize Monte Carlo simulations. 
The generated events are passed through the CDF detector simulation~\cite{Gerchtein:2003ba} and are reconstructed in the same way 
as the measured data. 

The simulated $t\bar{t}$ signal sample was generated with the \textsc{pythia} generator~\cite{Sjostrand:2000wi} using
a top quark mass of $m_t=178\;\mbox{GeV}/c^2$ which was the world average~\cite{Azzi:2004rc} of Run~I.
The values of $F_0$ and $F_+$  used in our standard model simulation are 0.7 and 0.0 respectively.
To check the assumption that neither the efficiency nor the resolution due to the reconstruction depend on
the values of $F_0$ and $F_+$, we use a customized version of the \textsc{herwig} Monte Carlo
program~\cite{Corcella:2000bw} in which the helicity of one
$W$ boson is fixed to be longitudinal, left-handed, or 
right-handed.  

%For checks we are using a customized version of the \textsc{herwig} Monte Carlo
%program~\cite{Corcella:2000bw} in which the helicity of one
%$W$ boson is fixed to be longitudinally polarized, left-handed or 
%right-handed.  

The selected $t\bar{t}$ candidate sample contains a certain level of background
contamination. Among the 82 observed events, we predict a background 
of $10.3 \pm 1.9$ events~\cite{Abulencia:2006in}.
The dominant sources are $W$ production in association with a quark-antiquark pair (31\%), $e.g.$ $\bar{q}q^\prime\rightarrow Wgg$
with $g\rightarrow b\bar{b}$ ($ c\bar{c}$) and
$g\rightarrow q''\bar{q}^{\prime\prime}$, ``mistagged'' events (24\%), in which a  
jet is erroneously tagged as a $b$-jet, and events where no $W$ boson (non-$W$ events) is produced (36\%),
$e.g.$ direct $b\bar{b}$ production with additional gluon radiation.
Additional sources are diboson ($WW$, $WZ$, $ZZ$) production (4.5\%)
and single-top production (4.5\%).
The {non-$W$} and mistag fractions are estimated using lepton trigger data.
The $W$ plus heavy flavor fraction is extracted using a sample of events simulated with \textsc{alpgen} \cite{Mangano:2002ea}.
The diboson and single-top rates are predicted based on their theoretical cross sections~\cite{Campbell:1999ah} and acceptances and efficiencies, which are derived from \textsc{pythia} and \textsc{madevent}~\cite{Maltoni:2002qb} 
simulations.

\section{Full Reconstruction of \boldmath${t\bar{t}}$ Pairs}
\label{sec:FullTTbar}

The measurement of $\cos\theta^*$ is based on fully reconstructing the top quarks through the four-momenta
of the decay products.
The challenge for the full reconstruction is to assign the observed jets to the decay products of the
hadronically decaying $W$ boson or the jets resulting from the $b$ quarks from the top-quark decays. All possible
assignments have to be considered. Thus, in each event there exist numerous hypotheses for the reconstruction of the $t\bar{t}$ pair. 
At the top quark reconstruction level, extra jet corrections are applied.
The calorimeter energy is corrected to correspond to the energy of the traversing particle, the
underlying event energy is subtracted, and, finally, the energy that is radiated outside the jet cone is added.
The $p_{\rm T}$ vector of the neutrino is derived from $\not\!\! \vec{E}_{\rm T}$.
To calculate the $z$-component of the neutrino momentum, a quadratic constraint using the $W\rightarrow \ell\nu$
decay kinematics is used, with the assumption that the $W$ boson mass equals the pole mass of 80.43~GeV$/c^2$.
If the solution of the equation is complex, the real part of the
solution is taken; otherwise the solution with the smaller value of $|p_{z,\nu}|$ is used.
Adding the resulting four-momentum of the neutrino and the four-momentum of the charged lepton leads to the 
correct $W$ boson four-vector in 78\% of simulated events.
In order to get all hypotheses for the semileptonically decaying top quark, we consider all combinations of the four-momentum of one of the 
selected jets and the four-momentum of the $W$ boson. 
The hadronically decaying $W$ boson is then reconstructed
by combining the four-momenta of two of the selected jets not assigned to the semileptonically decaying top quark. 
Adding the four-momenta of this $W$ boson and of one of the remaining jets results in the hadronically decaying top quark. 
This procedure leads to $\textstyle \frac{1}{2}$~$\cdot N_{\rm jets}!/(N_{\rm jets}-4)!$ different hypotheses for each event. 

For simulated events it is possible to determine the hypothesis which is closest to the true event.
This ``best hypothesis'' is defined as the hypothesis for which the deviation of the reconstructed top quarks and $W$ bosons 
from the generated particles in the $\eta$-$\phi$~plane is minimal.
Since this is not possible for measured data, we determine for each hypothesis a quantity $\Psi$ which gives a quantitative
estimate of how well this hypothesis matches the $t\bar{t}$ pair assumption, and we choose 
the hypothesis with the highest value of $\Psi$. 
Constraints on the mass of the hadronically decaying $W$ boson, on the mass difference between both reconstructed top quark masses,
on how $b$-like the jets assigned as $b$-jets in the $t\rightarrow Wb$ decays are, and on the reconstructed $E_{\rm T}$  
of the two top quarks enter the computation of $\Psi$.

%In order to choose one event interpretation in data, for each hypothesis, we determine a quantity $\Psi$, which gives a quantitative
%estimate of how well the hypothesis matches the $t\bar{t}$ pair assumption. 
%The quantities entering the computation of $\Psi$ are:
%\begin{enumerate}
%\item Constraints on the mass of the hadronically decaying $W$ boson and on
%the mass difference between both reconstructed top masses, which
%should be the same within the top mass width.
%\item The $b$ likeness of the $b$-jet candidates from the top pair.
%\item Constraint on the sum of the reconstructed transverse energy 
%of the two top quarks, which should, in leading order calculation, 
%be equal to the transverse energy of the event. 
%\end{enumerate}

We define $\Psi$ as\\
\begin{eqnarray}
  \Psi & = & \frac{1}{|\hat{f}_{E}-f_{E}| \cdot \chi^2} \cdot P_{\rm b} \;\; ,
  \label{eq:psi}
\end{eqnarray}
where $f_{E}$ is the sum of the transverse energies of the two top quarks divided by the total $E_{\rm T}$ of the event including $\not\!\! E_{\rm T}$:\\ 
\begin{eqnarray}
  f_{E} = \frac{\sqrt{p_{{\rm T},t \rightarrow b\ell\nu}^2 + m_{t \rightarrow b\ell\nu}^2} +
      \sqrt{p_{{\rm T},t \rightarrow bjj}^2 + m_{t \rightarrow bjj}^2}}
{\Sigma p_{\rm T,jet} + \not\!\! E_{\rm T} + E_{{\rm T},\,\ell}}\;\;,
  \label{eqn:p_energy}
\end{eqnarray}
where $p_{{\rm T},t \rightarrow b\ell\nu}$ and $p_{{\rm T},t \rightarrow bjj}$
are the reconstructed transverse momenta of the semileptonically and
hadronically decaying top quarks and $m_{t \rightarrow b\ell\nu}$ and $m_{t 
\rightarrow bjj}$ are the respective reconstructed top quark masses. 
The quantity $\Sigma p_{\rm T,jet}$ is the sum of 
the transverse momenta of the four jets used in the $t\bar{t}$ event hypothesis.
The transverse energy of the charged lepton is indicated with $E_{{\rm T},\,\ell}$.
The motivation for the definition of $f_{E}$ 
is that the $E_{\rm T}$ of the top quarks is approximately equal to the $E_{\rm T}$ of the entire event.
The mean value $\hat f_{E}$ of the $f_E$ distribution, obtained from the best hypothesis for each event of
a $t\bar{t}$ Monte Carlo simulation, is determined to be 1.014.

The quantity $\chi^2$ is defined as\\
\begin{eqnarray}
 \chi^2 = \frac{(m_{W\rightarrow jj}-\hat{m}_{W\rightarrow jj})^2}
 {\sigma_{m_{W\rightarrow jj}}^2} +
 \frac{(m_{t \rightarrow b\ell\nu} - m_{t \rightarrow bjj})^2}
 {\sigma_{\Delta m_{t}}^2}\;\; ,
\end{eqnarray}
where $m_{W\rightarrow jj}$ is the reconstructed mass of the hadronically
decaying $W$ boson and $m_{t \rightarrow b\ell\nu}$ and $m_{t \rightarrow bjj}$
are the reconstructed mass of the semileptonically decaying top quark and the
hadronically decaying top quark, respectively. The reconstructed mass of
the hadronically decaying $W$ boson should be equal to the mean value
$\hat m_{W\rightarrow jj}$ within the resolution $\sigma_{m_{W\rightarrow jj}}$ and the difference between both top quark masses should
be zero within the resolution $\sigma_{\Delta m_{t}}$. 
The values $\hat m_{W\rightarrow jj} = 79.5\, \mbox{GeV}/c^2$, $\sigma_{m_{W\rightarrow jj}} = 10.2\, \mbox{GeV}/c^2$, and $\sigma_{\Delta m_{t}} = 30.3\, \mbox{GeV}/c^2$ that we use are obtained from the corresponding mass 
distributions  using the best hypothesis of fully simulated $t\bar{t}$ events. 
The mass resolutions are dominated by the uncertainties in the jet energy reconstruction.
The jet energy scale is determined from dijet data events and simulated samples and checked using $\gamma$+jet and $Z$+jet 
events~\cite{Abulencia:2005aj,Abulencia:2005ak}. The value for $\hat m_{W\rightarrow jj}$ deviates from the measured $W$ boson pole mass $m_W=80.43\;\mbox{GeV}/c^2$.
The deviation is within the systematic uncertainties of the applied jet corrections. 

The quantity $P_{ b}$ is a measure of how $b$-like the two jets assigned as such by the event reconstruction are, and is defined as:
  \begin{eqnarray}
    P_{b} & = &(-\log{\cal P}_{t \rightarrow b\ell\nu}
-\log{\cal P}_{t \rightarrow bjj}) \cdot 10^{N_{\rm tag}} \;\; ,
  \end{eqnarray}
where ${\mathcal P}_{t \rightarrow b\, \ell\nu}$ and ${\mathcal P}_{t \rightarrow b\, jj}$ are the 
probabilities that the jets chosen to be the $b$-jets from the semileptonically 
and hadronically decaying top quark are consistent with the hypothesis of a light quark jet with zero lifetime.
This probability is calculated from the impact parameter of the tracks assigned to 
the jet in the $r$-$\phi$ plane~\cite{Abulencia:2006kv}.
The negative logarithm of that probability leads to large values for 
$b$-jets and small values for light flavor jets.
However, since a reconstructed secondary vertex is a stronger indication for $b$-jets than the probability based on the impact parameter,
the quantity $P_{b}$ should be given a higher weight when there are secondary vertex tagged jets. 
Since $-\log{\mathcal P}$ nearly always takes values smaller than 10, 
the logarithmic sum is multiplied by the factor $10^{N_{\rm tag}}$, where $N_{\rm tag}$ is the number of $b$-tagged jets (either 0, 1 or 2).

%After the full reconstruction, the four-vectors of the two top quarks are corrected. The momentum components remain 
%unchanged, while the energy is recalculated to force $m_t=178\;\mbox{GeV}/c^2$ 
%instead of the reconstructed mass. 

\begin{table}[tb]
\begin{center}
\caption[Quality of selected Hypothesis]{ Percentage of $t\bar{t}$ events that are reconstructed
within a particular $\sum\Delta R$, as defined in Eq.~\ref{eq:besthyp}.}
\label{tab:Cor-hyp}
\begin{tabular}{lc}
\hline \hline
                        & \multicolumn{1}{c}{fraction [\%]} \\ \hline
 best                   & 30.2 \\
\hline
 $\sum\Delta R < 1.5$   & 41.5 \\
 $\sum\Delta R < 3.0$   & 57.9 \\
 $\sum\Delta R < 4.5$   & 66.4 \\ \hline \hline
\end{tabular}
\end{center}
\end{table}

In order to estimate the quality of the criterion for choosing the most probable event
reconstruction based on the quantity $\Psi$, Monte Carlo studies are performed. 
We examine the sum of the distances in the $\eta$-$\phi$ plane associated with the semileptonically
decaying top quark ($\Delta R_{t\rightarrow b\ell\nu}$), the hadronically decaying top quark ($\Delta R_{t\rightarrow bjj}$), and the hadronically decaying $W$ boson ($\Delta R_{W\rightarrow jj}$). 

\begin{eqnarray}
  \sum\Delta R & = & \Delta R_{t\rightarrow b\ell\nu} +  \Delta R_{t\rightarrow bjj}
  + \Delta R_{W\rightarrow jj}.
\label{eq:besthyp}
\end{eqnarray}
The distance $\Delta R$ between a generated (``gen'') and a reconstructed (``rec'')  particle is given by 
$\Delta R = \sqrt{(\phi_{\rm gen}-\phi_{\rm rec})^2 + (\eta_{\rm gen}-\eta_{\rm rec})^2}$.	
Table \ref{tab:Cor-hyp} shows how often our selected hypothesis
has a value of $\sum\Delta R$ below a given value. We also state the fraction of events in which the chosen hypothesis is
the ``best hypothesis'' which is defined for each event as the hypothesis with the smallest value of $\sum \Delta R$.   

Our reconstruction method yields $\cos\theta^*$ resolutions comparable to other methods used in previous CDF measurements~\cite{Abulencia:2005aj}. In addition the present approach allows the inclusion of events with more than four jets in a consistent way.

\begin{figure}[t]
\begin{center}
\includegraphics[width=0.45\textwidth]{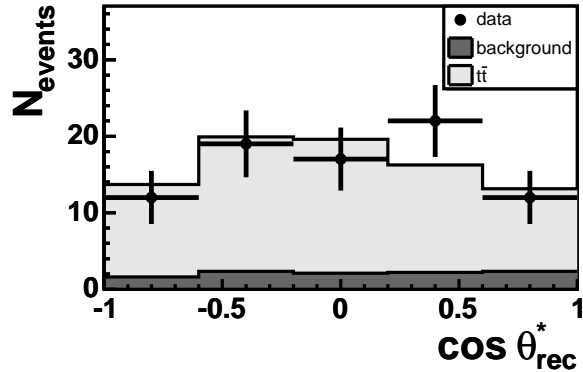}
\caption{\label{fig:costhetaData} Measured $\cos\theta^{*}_{\rm rec}$ distribution shown
together with the estimated signal and background. The $t\bar{t}$ signal events are modeled by the standard model Monte 
Carlo generator \textsc{pythia} with $m_{t} = 178$~GeV$/c^{2}$.}
  \end{center}	
\end{figure}	
Figure~\ref{fig:costhetaData} shows the distribution of the measured
$\cos\theta^*$ compared to the estimated signal and background distributions.

\section{Extraction of \boldmath${F_0}$ and ${F_+}$ and determination of the differential $t\bar{t}$ production cross section}
Since the number of events in the data set is small, we do not
simultaneously extract the fraction of longitudinally polarized
and right-handed $W$ bosons. 
We either fix $F_+$ to 0 and fit for $F_0$, or we fix $F_0$
to its expected value and fit for $F_+$.
Thus, only one free parameter is used in each fit.

To extract the single free parameter ($F_{0}$ or $F_+$), 
we use a binned maximum likelihood method.
The expected number of events in each bin is the sum of the expected background and signal. The latter is calculated
from the theoretical $\cos\theta^*$ distributions (Eq. \ref{cos_general})
for the three helicities of the $W$ boson. Integrating Eq.~\ref{cos_general} for each bin $i$ separately leads to a linear
 dependence of the expected number of signal events $\mu^{\rm sig}_i$ on $F_0$ and $F_+$:
%\begin{eqnarray}
%& F_0\cdot \underbrace{\int_{i_{min}}^{i_{max}}\left(\frac{3}{4}(1-\cos^{2}\theta^{*}) - \frac{3}{8}(1-\cos\theta^{*})^2 \right) d\cos\theta^*}_{A_i}\;\; +\nonumber\\[0.2cm]
%&  F_+\cdot \underbrace{\int_{i_{min}}^{i_{max}}\left(\frac{3}{8}(1+\cos\theta^{*})^2 
%- \frac{3}{8}(1-\cos\theta^{*})^2\right) d\cos\theta^*}_{B_i}\;\; +  \nonumber\\[0.2cm]
%& \underbrace{\int_{i_{min}}^{i_{max}}\frac{3}{8}(1-\cos\theta^{*})^2 d\cos\theta^*}_{C_i}\;\; ,
%\end{eqnarray}
\begin{eqnarray}
\mu^{\rm sig}_i \propto (1-F_0-F_+)\cdot f^-_i + (F_0) \cdot f^0_i + (F_+) \cdot f^+_i .
\end{eqnarray}
Here $f^0$, $f^-$ and $f^+$ are defined as: 
\begin{eqnarray}
 f_{i}^0 &=& \int_{a_{i}}^{b_{i}}\frac{3}{4}(1-\cos^{2}\theta^{*}) d\cos\theta^* ,\\
  f_{i}^- &=& \int_{a_{i}}^{b_{i}}\frac{3}{8}(1-\cos\theta^{*})^2 d\cos\theta^* ,\\ 
  f_{i}^+ &=& \int_{a_{i}}^{b_{i}}\frac{3}{8}(1+\cos\theta^{*})^2 d\cos\theta^* ,
\end{eqnarray}
where $a_{i}$ ($b_{i}$) is the lower (upper) edge of the i$^{th}$ bin. 

As mentioned above, the reconstruction of the $t\bar{t}$ process is not perfectly efficient.
Thus, in order to calculate the number of signal events $\mu^{\rm sig,obs}$ expected to be observed in a certain bin after the reconstruction, we consider acceptance and migration effects:  

\begin{equation}
\mu^{\rm sig,obs}_{k} \propto \sum_i \mu_{i}^{\rm sig} \cdot \epsilon_{i} \cdot S(i,k). 
\end{equation}

The migration matrix element $S(i,k)$ gives the probability for an event which was generated in bin $i$ to occur
in bin $k$ of the reconstructed $\cos\theta^*$ distribution.
Since the acceptance depends on $\cos\theta^{*}$, we weight the contribution of each bin $i$ with the efficiency $\epsilon_i$. 
Both $\epsilon_i$ and $S(i,k)$ are determined
using the standard model Monte Carlo generator \textsc{pythia}, assuming 
that $\epsilon_i$ and $S(i,k)$ are independent
of $F_0$ and $F_+$. This assumption has been verified using the customized \textsc{herwig}
samples described above, which have fixed $W$ helicities. 
%These samples have additionally been used
%to check whether the extraction method leads to the same $F_0$ or $F_+$ fractions,
%as contained in an arbitrary chosen mixture of the three samples for
%longitudinally polarized, left-handed and right-handed $W$ bosons. Good
%agreement between the simulated values and the extracted values has been
%observed.

With the number of expected events and the number of observed events in each bin, we minimize the negative logarithm of the
likelihood function by varying the free parameter $F_{0}$ or $F_{+}$. 

In addition, an upper limit for $F_+$ at the 95\% confidence
level (CL) is computed by integrating the likelihood function $L(F_+)$.
Since a Bayesian approach is pursued, we integrate only in the physical region $0\leq F_{+}\leq0.3$ applying a
prior distribution which is 1 in the interval [0,0.3] and 0 elsewhere.

In order to compare our observations with theory, the background estimate is subtracted from
the selected sample. To  correct for acceptance and reconstruction effects,
a transfer function $\tau$ is calculated.
The value $\tau_i$ for bin $i$ is the ratio of the normalized number of theoretically expected events and the normalized number
 of events after applying all selection cuts and performing the reconstruction.
For this calculation we use the fit result of $F_0$ or $F_+$.
Multiplying the background-subtracted number of events in bin $i$ with $\tau_i$ leads to the unfolded distribution.
Subsequently, this distribution is normalized to the $t\bar{t}$ production cross section of 
$\sigma_{t\bar{t}}~=~6.1\pm 0.9\,\mbox{pb}$ \cite{Cacciari:2003fi, Kidonakis:2003qe} assuming $m_{t} = 178$~GeV$/c^{2}$, which yields the desired
distribution of the differential cross section.

\section{Systematic Uncertainties}
\label{sec:syst}
The systematic uncertainties caused by theoretical modeling, detector effects, and the analysis method have been studied
using ensembles of simulated data samples.
Each sample is made up of signal and background events drawn from the respective templates.
The values for $F_{0}$ and $F_{+}$ are extracted using the same method as for the observed data sample. 
The systematic uncertainty for a certain source is then given by comparing the mean of the resulting $F_{0}$ and $F_{+}$ distributions
of the corresponding ensemble with the default values.

We account for possible bias from Monte Carlo modeling of $t\bar{t}$ events 
by comparing \textsc{herwig} and \textsc{pythia} event generators.

The contribution of the parton distribution function (PDF) uncertainty is determined by re-weighting the
$t\bar{t}$ events generated with {\sc cteq5l}~\cite{Lai:1999wy} for different sets of PDFs.
We add in quadrature the difference between {\sc mrst72} and {\sc mrst75}~\cite{Martin:1998sq} and between
the 20 pairs of {\sc cteq6m} eigenvectors.

To estimate the influence of initial-state and final-state radiation, we use templates from {\sc pythia} Monte Carlo simulations
in which the parameters for gluon radiation are varied to produce  
either less or more initial or final-state radiation~\cite{Abulencia:2005aj} compared to the standard setup.
The uncertainty due to the jet energy scale is
quantified by varying the jet energy scale within its $\pm 1 \sigma$ uncertainties~\cite{Bhatti:2005ai}.
We also investigate whether our method to choose one hypothesis for each single event contributes significantly to the total uncertainty.
Since the probable influence due to the $\chi^2$ and $f_E$ terms in the computation of the quantity $\Psi$ is already considered by
varying the jet energy scale, we study the impact of $P_{b}$ by omitting this term.
To estimate the contribution of the background rate uncertainty, we simultaneously add
or subtract, respectively, the values of one standard deviation 
of the estimated rates for the different processes.
The uncertainty due to the background shape uncertainty is
estimated by using each shape of the dominant three background
distributions alone instead of using a composite of  
these shapes. 

\begin{table}[t]
  \begin{center}
    \caption{Summary of systematic uncertainties. The total uncertainty is
    calculated by adding all the individual uncertainties in quadrature.}
    \begin{tabular}{lcccc}
       \hline\hline
              & \multicolumn{4}{c}{Uncertainties}\\
       Source & - $\Delta F_{0}$ & + $\Delta F_{0}$ & - $\Delta F_{+}$ & + $\Delta F_{+}$ \\ \hline
       Monte Carlo gen. & 0.022 & 0.022 & 0.010 & 0.010 \\
       Parton distribution functions        & 0.017 & 0.017 & 0.006 & 0.006 \\
       Initial-state radiation         & 0.010 & 0.010 & 0.007 & 0.007 \\ 
       Final-state radiation           & 0.005 & 0.005 & 0.002 & 0.002 \\ %\hline
       Jet energy scale                & 0.033 & 0.040 & 0.013 & 0.020 \\
       $b$-likeness of jet             & 0.009 & 0.009 & 0.008 & 0.008 \\
       Background normalization        & 0.002 & 0.004 & 0.000 & 0.003 \\ 
       Background shape                & 0.035 & 0.031 & 0.019 & 0.013 \\ %\hline     
      \bf Total               & \bf 0.057 & \bf 0.060 & \bf 0.028 & \bf 0.029 \\ \hline
       \hline
    \end{tabular}
    \label{tab:sys}
  \end{center}
\end{table}
The uncertainties are listed in Table \ref{tab:sys}. 
The largest contribution to the systematic uncertainty arises from
the jet energy-scale uncertainty, followed
by the uncertainty on the background shape.

Since the fraction of longitudinally polarized $W$ bosons depends explicitly on the
top quark mass, we do not include this dependence into the systematic uncertainties, but present our measurement assuming a certain top mass, namely 178 GeV$/c^2$.

\begin{figure*}[floatfix]
\begin{center}
\begin{minipage}{0.46\linewidth}
  \begin{center}
    \includegraphics[width=\textwidth]{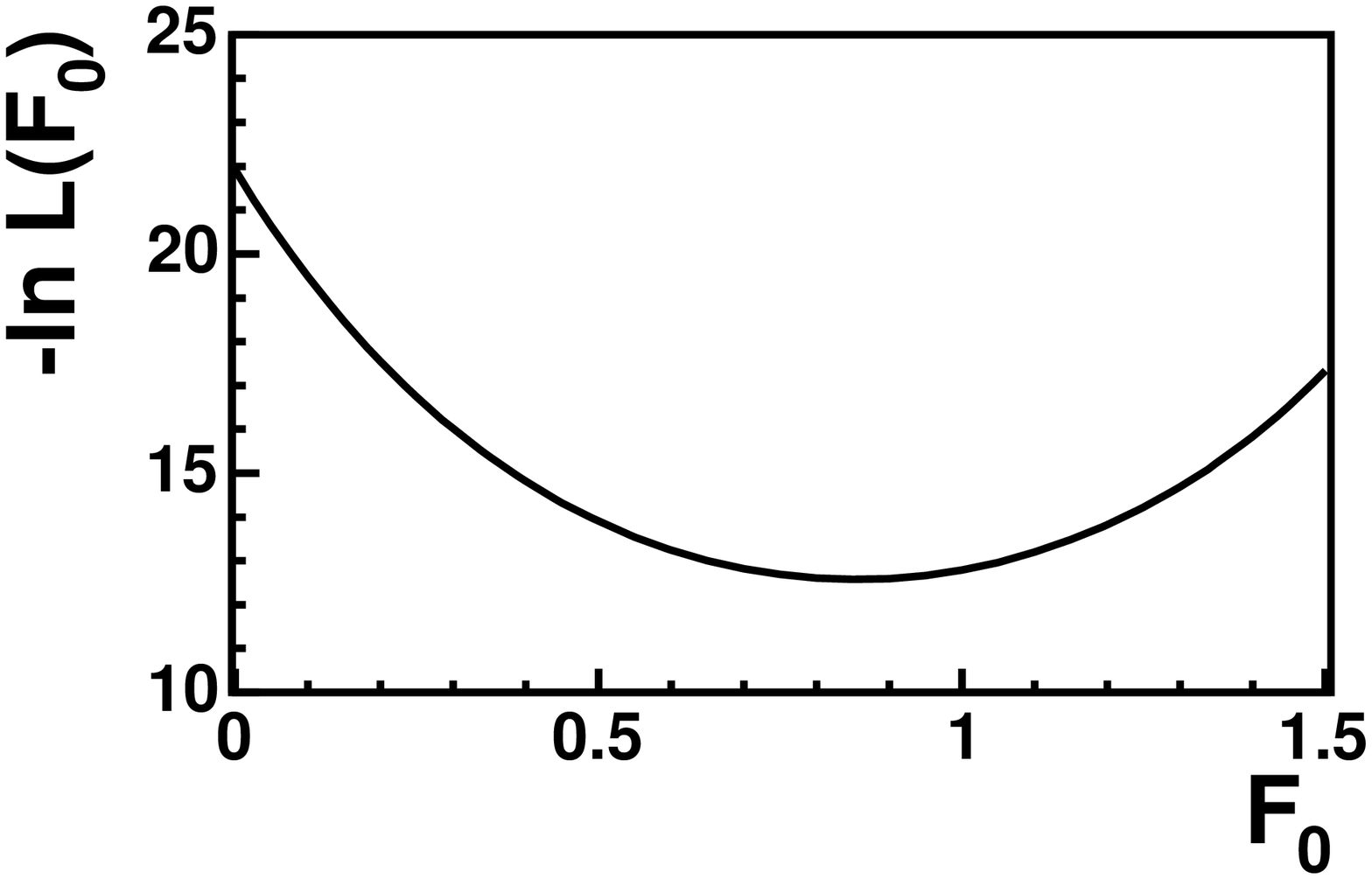} \\
    \includegraphics[width=\textwidth]{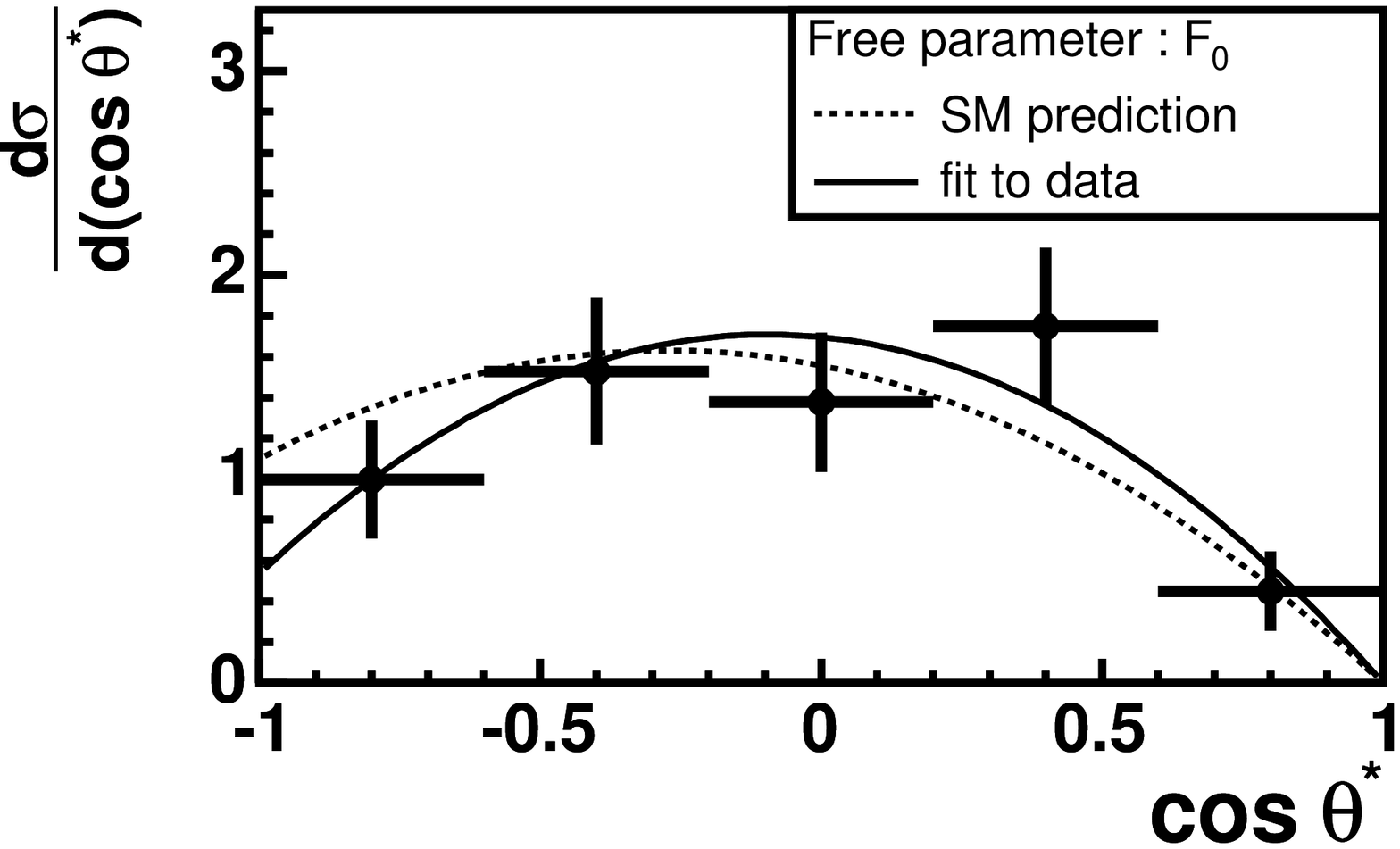}
  \end{center}
\end{minipage}
\hspace*{0.05\textwidth}
\begin{minipage}{0.46\linewidth}
  \begin{center}
    \includegraphics[width=\textwidth]{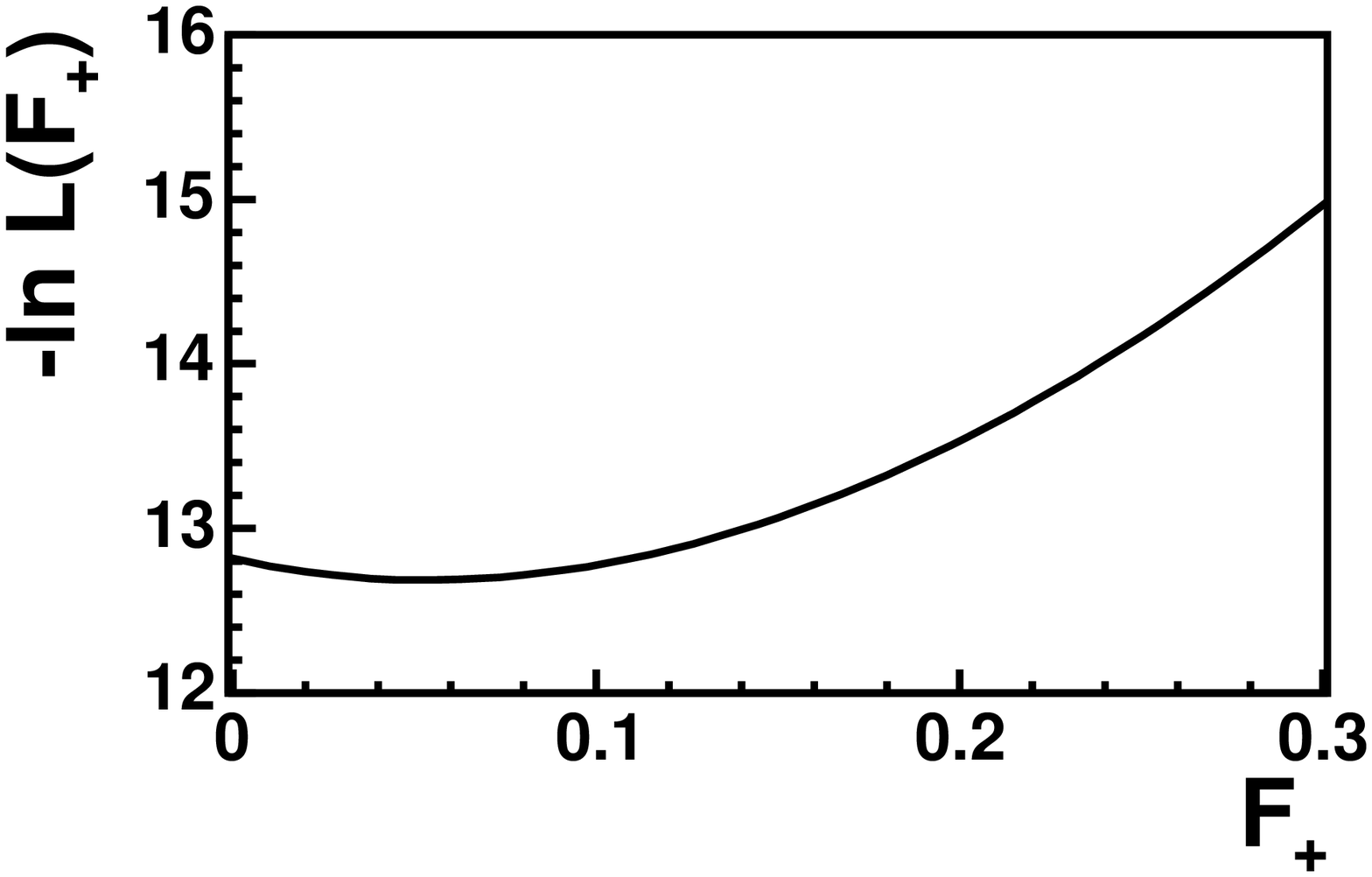} \\
    \includegraphics[width=\textwidth]{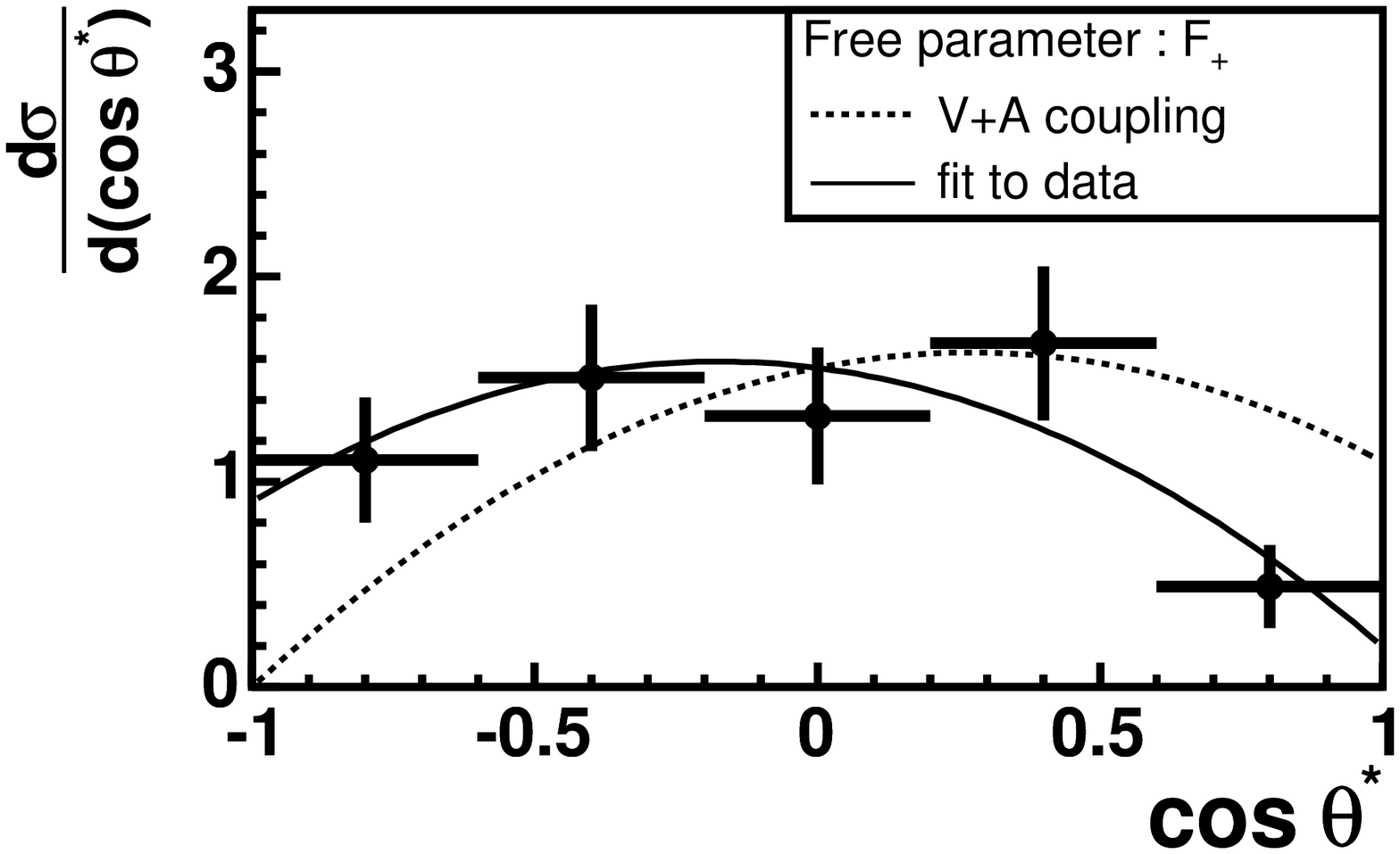}
  \end{center}
\end{minipage}
\begin{picture}(0,0)
  \unitlength1cm
  \put(-16.5,5.5) {\bf a)}
  \put(-16.5,0) {\bf c)}
  \put(-8,5.5) {\bf b)}
  \put(-8,0) {\bf d)}

\end{picture}
\caption{\label{fig:Wfit1} Extraction of the longitudinal ($F_0$)
and right-handed ($F_+$) fraction. For both fits $F_0$
and $F_+$ are used as single free parameter. In each case the other parameter is set to its
expected standard model value.
a,b) Negative log likelihood as a function
of $F_0$ or $F_+$.
c,d) Binned $\cos\theta^*$ distribution for data, corrected for acceptance and reconstruction effects.
The distributions corresponding to the fit results $F_{0} = 0.85$ and $F_{+} = 0.05$ are shown as a continuous functions. The dashed curve shows
the theoretical prediction for $F_0=0.7$ or for $F_+=0.3$.}
\end{center}
\end{figure*}

However, we investigate the dependence of the measured $F_{0}$ and $F_{+}$ on the top quark mass.  
For a shift of $+5$~GeV$/c^2$ ($-5$~GeV$/c^2$) in the top quark mass we estimate a deviation in $F_0$ of $+ 0.017 \pm 0.007$ ($- 0.017 \pm 0.007$),
which corresponds within the errors to the theoretical prediction $F_0=\frac{m_t^2}{2m_W^2+m_t^2}$.   
The standard model predicts a top mass independent value for $F_+$ of zero, whereas we see a small influence of
the top quark mass on our measurement of $F_+$. For a shift of $+5$ GeV$/c^2$ ($-5$ GeV$/c^2$) in the top quark mass we estimate a deviation in $F_+$ of $+ 0.008 \pm 0.003$ ($- 0.008 \pm 0.003$).

\section{Results}
\label{sec:results}

We have presented a method for the measurement of the fractions $F_0$ and $F_+$ of longitudinally
polarized and right-handed $W$ bosons in top quark decays
using a selected data sample with an integrated luminosity of approximately $318\;\mbox{pb}^{-1}$ 
collected with the CDF II detector.

Taking the systematic uncertainties into account, assuming a top quark mass of $m_t = 178$~GeV/$c^2$, 
and assuming that the non-measured fraction is equal to
the standard model expectation, the final result 
for the fractions of longitudinally polarized
and right-handed $W$ bosons is
\begin{eqnarray*}
F_0= 0.85^{+0.15}_{-0.22}\;(\mathrm{stat})\,\pm 0.06\,(\mathrm{syst}) ,\\ 
F_+= 0.05^{+0.11}_{-0.05}\;(\mathrm{stat})\,\pm 0.03\,(\mathrm{syst}) .
\end{eqnarray*}

We obtain an upper limit on the fraction of right-handed $W$ bosons of $F_+\le 0.26$ at the 95\% CL. 
The systematic uncertainties are incorporated by convoluting $L(F_+)$ with a Gaussian with a mean of zero and a width equal to the total systematic uncertainty.
 
Figure~\ref{fig:Wfit1}a (b) shows the negative log-likelihood as a 
function of $F_{0}$ ($F_+$), where the minimum represents
the result of the fit.
Our method provides the possibility to correct the distribution of observed $\cos\theta^*$ for the selected sample for acceptance and reconstruction effects.
Figures~\ref{fig:Wfit1}c and~\ref{fig:Wfit1}d show the unfolded distribution, normalized to the theoretical $t\bar{t}$ cross section, in comparison with
theoretical predictions for standard model and a $V+A$ model in the case of $F_{+}$.
As one can see, the observation is compatible with the  standard model prediction. Also the measured values for $F_{0}$ and $F_{+}$
are in good agreement with the standard model.

\section{Acknowledgments}

We thank the Fermilab staff and the technical staffs of the participating institutions for their vital contributions. This work was supported by the U.S. Department of Energy and National Science Foundation; the Italian Istituto Nazionale di Fisica Nucleare; the Ministry of Education, Culture, Sports, Science and Technology of Japan; the Natural Sciences and Engineering Research Council of Canada; the National Science Council of the Republic of China; the Swiss National Science Foundation; the A.P. Sloan Foundation; the Research Corporation; the Bundesministerium f\"ur Bildung und Forschung, Germany; the Korean Science and Engineering Foundation and the Korean Research Foundation; the Particle Physics and Astronomy Research Council and the Royal Society, UK; the Institut National de Physique Nucleaire et Physique des Particules/CNRS; the Russian Foundation for Basic Research; the Comisi\'on Interministerial de Ciencia y Tecnolog\'{\i}a, Spain; the European Community's Human Potential Programme under contract HPRN-CT-2002-00292; and the Academy of Finland.

\bibliography{bibiWhel}

\begin{thebibliography}{53}
\expandafter\ifx\csname natexlab\endcsname\relax\def\natexlab#1{#1}\fi
\expandafter\ifx\csname bibnamefont\endcsname\relax
  \def\bibnamefont#1{#1}\fi
\expandafter\ifx\csname bibfnamefont\endcsname\relax
  \def\bibfnamefont#1{#1}\fi
\expandafter\ifx\csname citenamefont\endcsname\relax
  \def\citenamefont#1{#1}\fi
\expandafter\ifx\csname url\endcsname\relax
  \def\url#1{\texttt{#1}}\fi
\expandafter\ifx\csname urlprefix\endcsname\relax\def\urlprefix{URL }\fi
\providecommand{\bibinfo}[2]{#2}
\providecommand{\eprint}[2][]{\url{#2}}

\bibitem[{\citenamefont{Abe \textit{et~al.}}(1995)}]{Abe:1995hr}
\bibinfo{author}{\bibfnamefont{F.}~\bibnamefont{Abe}}
  \bibnamefont{\textit{et~al.}} (\bibinfo{collaboration}{CDF~Collaboration}),
  \bibinfo{journal}{Phys. Rev. Lett.} \textbf{\bibinfo{volume}{74}},
  \bibinfo{pages}{2626} (\bibinfo{year}{1995}).

\bibitem[{\citenamefont{Abachi \textit{et~al.}}(1995)}]{Abachi:1995iq}
\bibinfo{author}{\bibfnamefont{S.}~\bibnamefont{Abachi}}
  \bibnamefont{\textit{et~al.}} (\bibinfo{collaboration}{D\O~Collaboration}),
  \bibinfo{journal}{Phys. Rev. Lett.} \textbf{\bibinfo{volume}{74}},
  \bibinfo{pages}{2632} (\bibinfo{year}{1995}).

\bibitem[{\citenamefont{Group}(2006)}]{Group:2006qt}
\bibinfo{author}{\bibfnamefont{T.~E.~W.} \bibnamefont{Group}}
  (\bibinfo{collaboration}{Tevatron Electroweak Working Group})
  (\bibinfo{year}{2006}), \eprint{hep-ex/0603039}.

\bibitem[{\citenamefont{Abulencia
  \textit{et~al.}}(2006{\natexlab{a}})}]{Abulencia:2005ak}
\bibinfo{author}{\bibfnamefont{A.}~\bibnamefont{Abulencia}}
  \bibnamefont{\textit{et~al.}} (\bibinfo{collaboration}{CDF~Collaboration}),
  \bibinfo{journal}{Phys. Rev. Lett.} \textbf{\bibinfo{volume}{96}},
  \bibinfo{pages}{022004} (\bibinfo{year}{2006}{\natexlab{a}}).

\bibitem[{\citenamefont{Cornwall \textit{et~al.}}(1974)\citenamefont{Cornwall,
  Levin, and Tiktopoulos}}]{Cornwall:1974km}
\bibinfo{author}{\bibfnamefont{J.~M.} \bibnamefont{Cornwall}},
  \bibinfo{author}{\bibfnamefont{D.~N.} \bibnamefont{Levin}}, \bibnamefont{and}
  \bibinfo{author}{\bibfnamefont{G.}~\bibnamefont{Tiktopoulos}},
  \bibinfo{journal}{Phys. Rev. D} \textbf{\bibinfo{volume}{10}},
  \bibinfo{pages}{1145} (\bibinfo{year}{1974}).

\bibitem[{\citenamefont{Lee \textit{et~al.}}(1977)\citenamefont{Lee, Quigg, and
  Thacker}}]{Lee:1977eg}
\bibinfo{author}{\bibfnamefont{B.~W.} \bibnamefont{Lee}},
  \bibinfo{author}{\bibfnamefont{C.}~\bibnamefont{Quigg}}, \bibnamefont{and}
  \bibinfo{author}{\bibfnamefont{H.~B.} \bibnamefont{Thacker}},
  \bibinfo{journal}{Phys. Rev. D} \textbf{\bibinfo{volume}{16}},
  \bibinfo{pages}{1519} (\bibinfo{year}{1977}).

\bibitem[{\citenamefont{K\"uhn}(1996)}]{Kuhn:1996ug}
\bibinfo{author}{\bibfnamefont{J.~H.} \bibnamefont{K\"uhn}}
  (\bibinfo{year}{1996}), Lectures delivered at 23rd SLAC Summer Institute,
  \eprint{hep-ph/9707321}.

\bibitem[{\citenamefont{Kane \textit{et~al.}}(1992)\citenamefont{Kane,
  Ladinsky, and Yuan}}]{Kane:1991bg}
\bibinfo{author}{\bibfnamefont{G.~L.} \bibnamefont{Kane}},
  \bibinfo{author}{\bibfnamefont{G.~A.} \bibnamefont{Ladinsky}},
  \bibnamefont{and} \bibinfo{author}{\bibfnamefont{C.~P.} \bibnamefont{Yuan}},
  \bibinfo{journal}{Phys. Rev. D} \textbf{\bibinfo{volume}{45}},
  \bibinfo{pages}{124} (\bibinfo{year}{1992}).

\bibitem[{\citenamefont{Eidelman \textit{et~al.}}(2004)}]{Eidelman:2004wy}
\bibinfo{author}{\bibfnamefont{S.}~\bibnamefont{Eidelman}}
  \bibnamefont{\textit{et~al.}} (\bibinfo{collaboration}{Particle Data Group}),
  \bibinfo{journal}{Phys. Lett. B} \textbf{\bibinfo{volume}{592}}
  (\bibinfo{year}{2004}).

\bibitem[{\citenamefont{Jezabek and K\"uhn}(1989)}]{jezabekKuehn1989}
\bibinfo{author}{\bibfnamefont{M.}~\bibnamefont{Jezabek}} \bibnamefont{and}
  \bibinfo{author}{\bibfnamefont{J.~H.} \bibnamefont{K\"uhn}},
  \bibinfo{journal}{Nucl. Phys.} \textbf{\bibinfo{volume}{B 314}},
  \bibinfo{pages}{1} (\bibinfo{year}{1989}).

\bibitem[{\citenamefont{Jezabek and K\"uhn}(1988)}]{jezabekKuehn1988}
\bibinfo{author}{\bibfnamefont{M.}~\bibnamefont{Jezabek}} \bibnamefont{and}
  \bibinfo{author}{\bibfnamefont{J.~H.} \bibnamefont{K\"uhn}},
  \bibinfo{journal}{Phys. Lett. B} \textbf{\bibinfo{volume}{207}},
  \bibinfo{pages}{91} (\bibinfo{year}{1988}).

\bibitem[{\citenamefont{Czarnecki and Melnikow}(1999)}]{czarnecki1999}
\bibinfo{author}{\bibfnamefont{A.}~\bibnamefont{Czarnecki}} \bibnamefont{and}
  \bibinfo{author}{\bibfnamefont{K.}~\bibnamefont{Melnikow}},
  \bibinfo{journal}{Nucl. Phys.} \textbf{\bibinfo{volume}{B 544}},
  \bibinfo{pages}{520} (\bibinfo{year}{1999}).

\bibitem[{\citenamefont{Chetyrkin
  \textit{et~al.}}(1999)\citenamefont{Chetyrkin, Harlander, Seidensticker, and
  Steinhauser}}]{chetyrkin1999}
\bibinfo{author}{\bibfnamefont{K.~G.} \bibnamefont{Chetyrkin}},
  \bibinfo{author}{\bibfnamefont{R.}~\bibnamefont{Harlander}},
  \bibinfo{author}{\bibfnamefont{T.}~\bibnamefont{Seidensticker}},
  \bibnamefont{and}
  \bibinfo{author}{\bibfnamefont{M.}~\bibnamefont{Steinhauser}},
  \bibinfo{journal}{Phys. Rev. D} \textbf{\bibinfo{volume}{60}},
  \bibinfo{pages}{114015} (\bibinfo{year}{1999}).

\bibitem[{\citenamefont{Denner and Sack}(1991)}]{denner1991}
\bibinfo{author}{\bibfnamefont{A.}~\bibnamefont{Denner}} \bibnamefont{and}
  \bibinfo{author}{\bibfnamefont{T.}~\bibnamefont{Sack}},
  \bibinfo{journal}{Nucl. Phys.} \textbf{\bibinfo{volume}{B 358}},
  \bibinfo{pages}{46} (\bibinfo{year}{1991}).

\bibitem[{\citenamefont{Migneron \textit{et~al.}}(1991)\citenamefont{Migneron,
  Eilam, Mendel, and Soni}}]{migneron1991}
\bibinfo{author}{\bibfnamefont{R.}~\bibnamefont{Migneron}},
  \bibinfo{author}{\bibfnamefont{G.}~\bibnamefont{Eilam}},
  \bibinfo{author}{\bibfnamefont{R.~R.} \bibnamefont{Mendel}},
  \bibnamefont{and} \bibinfo{author}{\bibfnamefont{A.}~\bibnamefont{Soni}},
  \bibinfo{journal}{Phys. Rev. Lett.} \textbf{\bibinfo{volume}{66}},
  \bibinfo{pages}{3105} (\bibinfo{year}{1991}).

\bibitem[{\citenamefont{Fischer \textit{et~al.}}(1999)\citenamefont{Fischer,
  Groote, Korner, Mauser, and Lampe}}]{Fischer:1998gs}
\bibinfo{author}{\bibfnamefont{M.}~\bibnamefont{Fischer}},
  \bibinfo{author}{\bibfnamefont{S.}~\bibnamefont{Groote}},
  \bibinfo{author}{\bibfnamefont{J.~G.} \bibnamefont{Korner}},
  \bibinfo{author}{\bibfnamefont{M.~C.} \bibnamefont{Mauser}},
  \bibnamefont{and} \bibinfo{author}{\bibfnamefont{B.}~\bibnamefont{Lampe}},
  \bibinfo{journal}{Phys. Lett. B} \textbf{\bibinfo{volume}{451}},
  \bibinfo{pages}{406} (\bibinfo{year}{1999}).

\bibitem[{\citenamefont{Fischer \textit{et~al.}}(2001)\citenamefont{Fischer,
  Groote, Korner, and Mauser}}]{Fischer:2000kx}
\bibinfo{author}{\bibfnamefont{M.}~\bibnamefont{Fischer}},
  \bibinfo{author}{\bibfnamefont{S.}~\bibnamefont{Groote}},
  \bibinfo{author}{\bibfnamefont{J.~G.} \bibnamefont{Korner}},
  \bibnamefont{and} \bibinfo{author}{\bibfnamefont{M.~C.}
  \bibnamefont{Mauser}}, \bibinfo{journal}{Phys. Rev. D}
  \textbf{\bibinfo{volume}{63}}, \bibinfo{pages}{031501}
  (\bibinfo{year}{2001}).

\bibitem[{\citenamefont{Fischer \textit{et~al.}}(2002)\citenamefont{Fischer,
  Groote, Korner, and Mauser}}]{Fischer:2001gp}
\bibinfo{author}{\bibfnamefont{M.}~\bibnamefont{Fischer}},
  \bibinfo{author}{\bibfnamefont{S.}~\bibnamefont{Groote}},
  \bibinfo{author}{\bibfnamefont{J.~G.} \bibnamefont{Korner}},
  \bibnamefont{and} \bibinfo{author}{\bibfnamefont{M.~C.}
  \bibnamefont{Mauser}}, \bibinfo{journal}{Phys. Rev. D}
  \textbf{\bibinfo{volume}{65}}, \bibinfo{pages}{054036}
  (\bibinfo{year}{2002}).

\bibitem[{\citenamefont{Do \textit{et~al.}}(2003)\citenamefont{Do, Groote,
  Korner, and Mauser}}]{Do:2002ky}
\bibinfo{author}{\bibfnamefont{H.~S.} \bibnamefont{Do}},
  \bibinfo{author}{\bibfnamefont{S.}~\bibnamefont{Groote}},
  \bibinfo{author}{\bibfnamefont{J.~G.} \bibnamefont{Korner}},
  \bibnamefont{and} \bibinfo{author}{\bibfnamefont{M.~C.}
  \bibnamefont{Mauser}}, \bibinfo{journal}{Phys. Rev. D}
  \textbf{\bibinfo{volume}{67}}, \bibinfo{pages}{091501}
  (\bibinfo{year}{2003}).

\bibitem[{\citenamefont{Peccei \textit{et~al.}}(1991)\citenamefont{Peccei,
  Peris, and Zhang}}]{Peccei:1990uv}
\bibinfo{author}{\bibfnamefont{R.~D.} \bibnamefont{Peccei}},
  \bibinfo{author}{\bibfnamefont{S.}~\bibnamefont{Peris}}, \bibnamefont{and}
  \bibinfo{author}{\bibfnamefont{X.}~\bibnamefont{Zhang}},
  \bibinfo{journal}{Nucl. Phys.} \textbf{\bibinfo{volume}{B 349}},
  \bibinfo{pages}{305} (\bibinfo{year}{1991}).

\bibitem[{\citenamefont{Wang \textit{et~al.}}(2005)\citenamefont{Wang, Zhang,
  and Qiao}}]{Wang:2005ra}
\bibinfo{author}{\bibfnamefont{X.-L.} \bibnamefont{Wang}},
  \bibinfo{author}{\bibfnamefont{Q.-L.} \bibnamefont{Zhang}}, \bibnamefont{and}
  \bibinfo{author}{\bibfnamefont{Q.-P.} \bibnamefont{Qiao}},
  \bibinfo{journal}{Phys. Rev. D} \textbf{\bibinfo{volume}{71}},
  \bibinfo{pages}{014035} (\bibinfo{year}{2005}).

\bibitem[{\citenamefont{Chen \textit{et~al.}}(2005)\citenamefont{Chen, Larios,
  and Yuan}}]{Chen:2005vr}
\bibinfo{author}{\bibfnamefont{C.-R.} \bibnamefont{Chen}},
  \bibinfo{author}{\bibfnamefont{F.}~\bibnamefont{Larios}}, \bibnamefont{and}
  \bibinfo{author}{\bibfnamefont{C.~P.} \bibnamefont{Yuan}},
  \bibinfo{journal}{Phys. Lett. B} \textbf{\bibinfo{volume}{631}},
  \bibinfo{pages}{126} (\bibinfo{year}{2005}).

\bibitem[{\citenamefont{Abulencia
  \textit{et~al.}}(2006{\natexlab{b}})}]{Abulencia:2006in}
\bibinfo{author}{\bibfnamefont{A.}~\bibnamefont{Abulencia}}
  \bibnamefont{\textit{et~al.}} (\bibinfo{collaboration}{CDF~Collaboration}),
  \bibinfo{journal}{Phys. Rev. Lett.} \textbf{\bibinfo{volume}{97}},
  \bibinfo{pages}{082004} (\bibinfo{year}{2006}{\natexlab{b}}).

\bibitem[{\citenamefont{Acosta
  \textit{et~al.}}(2005{\natexlab{a}})}]{Acosta:2004mb}
\bibinfo{author}{\bibfnamefont{D.}~\bibnamefont{Acosta}}
  \bibnamefont{\textit{et~al.}} (\bibinfo{collaboration}{CDF~Collaboration}),
  \bibinfo{journal}{Phys. Rev. D} \textbf{\bibinfo{volume}{71}},
  \bibinfo{pages}{031101} (\bibinfo{year}{2005}{\natexlab{a}}).

\bibitem[{\citenamefont{Abulencia
  \textit{et~al.}}(2006{\natexlab{c}})}]{Abulencia:2005xf}
\bibinfo{author}{\bibfnamefont{A.}~\bibnamefont{Abulencia}}
  \bibnamefont{\textit{et~al.}} (\bibinfo{collaboration}{CDF~Collaboration}),
  \bibinfo{journal}{Phys. Rev. D} \textbf{\bibinfo{volume}{73}},
  \bibinfo{pages}{111103} (\bibinfo{year}{2006}{\natexlab{c}}).

\bibitem[{\citenamefont{Abulencia}(2006)}]{Abulencia:2006iy}
\bibinfo{author}{\bibfnamefont{A.}~\bibnamefont{Abulencia}}
  (\bibinfo{year}{2006}), submitted to Phys. Rev. Lett.,
  \eprint{hep-ex/0608062}.

\bibitem[{\citenamefont{Affolder
  \textit{et~al.}}(2000{\natexlab{a}})}]{Affolder:1999mp}
\bibinfo{author}{\bibfnamefont{A.}~\bibnamefont{Affolder}}
  \bibnamefont{\textit{et~al.}} (\bibinfo{collaboration}{CDF~Collaboration}),
  \bibinfo{journal}{Phys. Rev. Lett.} \textbf{\bibinfo{volume}{84}},
  \bibinfo{pages}{216} (\bibinfo{year}{2000}{\natexlab{a}}).

\bibitem[{\citenamefont{Abazov
  \textit{et~al.}}(2005{\natexlab{a}})}]{Abazov:2004ym}
\bibinfo{author}{\bibfnamefont{V.~M.} \bibnamefont{Abazov}}
  \bibnamefont{\textit{et~al.}} (\bibinfo{collaboration}{D\O~Collaboration}),
  \bibinfo{journal}{Phys. Lett. B} \textbf{\bibinfo{volume}{617}},
  \bibinfo{pages}{1} (\bibinfo{year}{2005}{\natexlab{a}}).

\bibitem[{\citenamefont{Abazov
  \textit{et~al.}}(2005{\natexlab{b}})}]{Abazov:2005fk}
\bibinfo{author}{\bibfnamefont{V.~M.} \bibnamefont{Abazov}}
  \bibnamefont{\textit{et~al.}} (\bibinfo{collaboration}{D\O~Collaboration}),
  \bibinfo{journal}{Phys. Rev. D} \textbf{\bibinfo{volume}{72}},
  \bibinfo{pages}{011104} (\bibinfo{year}{2005}{\natexlab{b}}).

\bibitem[{\citenamefont{Acosta
  \textit{et~al.}}(2005{\natexlab{b}})}]{Acosta:2004yw}
\bibinfo{author}{\bibfnamefont{D.}~\bibnamefont{Acosta}}
  \bibnamefont{\textit{et~al.}} (\bibinfo{collaboration}{CDF~Collaboration}),
  \bibinfo{journal}{Phys. Rev. D} \textbf{\bibinfo{volume}{71}},
  \bibinfo{pages}{032001} (\bibinfo{year}{2005}{\natexlab{b}}).

\bibitem[{\citenamefont{Affolder \textit{et~al.}}(2004)}]{Affolder:2003ep}
\bibinfo{author}{\bibfnamefont{A.}~\bibnamefont{Affolder}}
  \bibnamefont{\textit{et~al.}} (\bibinfo{collaboration}{CDF~Collaboration}),
  \bibinfo{journal}{Nucl. Instrum. Methods A} \textbf{\bibinfo{volume}{526}},
  \bibinfo{pages}{249} (\bibinfo{year}{2004}).

\bibitem[{\citenamefont{Hill}(2004)}]{Hill:2004qb}
\bibinfo{author}{\bibfnamefont{C.~S.} \bibnamefont{Hill}}
  (\bibinfo{collaboration}{CDF~Collaboration}), \bibinfo{journal}{Nucl.
  Instrum. Methods A} \textbf{\bibinfo{volume}{530}}, \bibinfo{pages}{1}
  (\bibinfo{year}{2004}).

\bibitem[{\citenamefont{Sill}(2000)}]{Sill:2000zz}
\bibinfo{author}{\bibfnamefont{A.}~\bibnamefont{Sill}}
  (\bibinfo{collaboration}{CDF~Collaboration}), \bibinfo{journal}{Nucl.
  Instrum. Methods A} \textbf{\bibinfo{volume}{447}}, \bibinfo{pages}{1}
  (\bibinfo{year}{2000}).

\bibitem[{\citenamefont{Affolder
  \textit{et~al.}}(2000{\natexlab{b}})}]{Affolder:2000tj}
\bibinfo{author}{\bibfnamefont{A.}~\bibnamefont{Affolder}}
  \bibnamefont{\textit{et~al.}} (\bibinfo{collaboration}{CDF~Collaboration}),
  \bibinfo{journal}{Nucl. Instrum. Methods A} \textbf{\bibinfo{volume}{453}},
  \bibinfo{pages}{84} (\bibinfo{year}{2000}{\natexlab{b}}).

\bibitem[{\citenamefont{Balka \textit{et~al.}}(1988)}]{Balka:1987ty}
\bibinfo{author}{\bibfnamefont{L.}~\bibnamefont{Balka}}
  \bibnamefont{\textit{et~al.}} (\bibinfo{collaboration}{CDF~Collaboration}),
  \bibinfo{journal}{Nucl. Instrum. Methods A} \textbf{\bibinfo{volume}{267}},
  \bibinfo{pages}{272} (\bibinfo{year}{1988}).

\bibitem[{\citenamefont{Bertolucci \textit{et~al.}}(1988)}]{Bertolucci:1987zn}
\bibinfo{author}{\bibfnamefont{S.}~\bibnamefont{Bertolucci}}
  \bibnamefont{\textit{et~al.}} (\bibinfo{collaboration}{CDF~Collaboration}),
  \bibinfo{journal}{Nucl. Instrum. Methods A} \textbf{\bibinfo{volume}{267}},
  \bibinfo{pages}{301} (\bibinfo{year}{1988}).

\bibitem[{\citenamefont{Albrow \textit{et~al.}}(2002)}]{Albrow:2001jw}
\bibinfo{author}{\bibfnamefont{M.~G.} \bibnamefont{Albrow}}
  \bibnamefont{\textit{et~al.}} (\bibinfo{collaboration}{CDF~Collaboration}),
  \bibinfo{journal}{Nucl. Instrum. Methods A} \textbf{\bibinfo{volume}{480}},
  \bibinfo{pages}{524} (\bibinfo{year}{2002}).

\bibitem[{\citenamefont{Ascoli \textit{et~al.}}(1988)}]{Ascoli:1987av}
\bibinfo{author}{\bibfnamefont{G.}~\bibnamefont{Ascoli}}
  \bibnamefont{\textit{et~al.}}, \bibinfo{journal}{Nucl. Instrum. Methods A}
  \textbf{\bibinfo{volume}{268}}, \bibinfo{pages}{33} (\bibinfo{year}{1988}).

\bibitem[{\citenamefont{Bhatti \textit{et~al.}}(2006)}]{Bhatti:2005ai}
\bibinfo{author}{\bibfnamefont{A.}~\bibnamefont{Bhatti}}
  \bibnamefont{\textit{et~al.}}, \bibinfo{journal}{Nucl. Instrum. Methods A}
  \textbf{\bibinfo{volume}{566}}, \bibinfo{pages}{375} (\bibinfo{year}{2006}).

\bibitem[{\citenamefont{Acosta
  \textit{et~al.}}(2005{\natexlab{c}})}]{Acosta:2004hw}
\bibinfo{author}{\bibfnamefont{D.}~\bibnamefont{Acosta}}
  \bibnamefont{\textit{et~al.}} (\bibinfo{collaboration}{CDF~Collaboration}),
  \bibinfo{journal}{Phys. Rev. D} \textbf{\bibinfo{volume}{71}},
  \bibinfo{pages}{052003} (\bibinfo{year}{2005}{\natexlab{c}}).

\bibitem[{\citenamefont{Gerchtein and Paulini}(2003)}]{Gerchtein:2003ba}
\bibinfo{author}{\bibfnamefont{E.}~\bibnamefont{Gerchtein}} \bibnamefont{and}
  \bibinfo{author}{\bibfnamefont{M.}~\bibnamefont{Paulini}}
  (\bibinfo{year}{2003}), Talk given at 2003 Conference on Computing in
  High-Energy and Nuclear Physics, \eprint{physics/0306031}.

\bibitem[{\citenamefont{Sjostrand \textit{et~al.}}(2001)}]{Sjostrand:2000wi}
\bibinfo{author}{\bibfnamefont{T.}~\bibnamefont{Sjostrand}}
  \bibnamefont{\textit{et~al.}}, \bibinfo{journal}{Comput. Phys. Commun.}
  \textbf{\bibinfo{volume}{135}}, \bibinfo{pages}{238} (\bibinfo{year}{2001}).

\bibitem[{\citenamefont{Azzi \textit{et~al.}}(2004)}]{Azzi:2004rc}
\bibinfo{author}{\bibfnamefont{P.}~\bibnamefont{Azzi}}
  \bibnamefont{\textit{et~al.}} (\bibinfo{collaboration}{CDF~Collaboration})
  (\bibinfo{year}{2004}), \eprint{hep-ex/0404010}.

\bibitem[{\citenamefont{Corcella \textit{et~al.}}(2001)}]{Corcella:2000bw}
\bibinfo{author}{\bibfnamefont{G.}~\bibnamefont{Corcella}}
  \bibnamefont{\textit{et~al.}}, \bibinfo{journal}{J. High Energy Phys.}
  \textbf{\bibinfo{volume}{01}}, \bibinfo{pages}{010} (\bibinfo{year}{2001}).

\bibitem[{\citenamefont{Mangano \textit{et~al.}}(2003)\citenamefont{Mangano,
  Moretti, Piccinini, Pittau, and Polosa}}]{Mangano:2002ea}
\bibinfo{author}{\bibfnamefont{M.~L.} \bibnamefont{Mangano}},
  \bibinfo{author}{\bibfnamefont{M.}~\bibnamefont{Moretti}},
  \bibinfo{author}{\bibfnamefont{F.}~\bibnamefont{Piccinini}},
  \bibinfo{author}{\bibfnamefont{R.}~\bibnamefont{Pittau}}, \bibnamefont{and}
  \bibinfo{author}{\bibfnamefont{A.~D.} \bibnamefont{Polosa}},
  \bibinfo{journal}{J. High Energy Phys.} \textbf{\bibinfo{volume}{07}},
  \bibinfo{pages}{001} (\bibinfo{year}{2003}).

\bibitem[{\citenamefont{Campbell and Ellis}(1999)}]{Campbell:1999ah}
\bibinfo{author}{\bibfnamefont{J.~M.} \bibnamefont{Campbell}} \bibnamefont{and}
  \bibinfo{author}{\bibfnamefont{R.~K.} \bibnamefont{Ellis}},
  \bibinfo{journal}{Phys. Rev. D} \textbf{\bibinfo{volume}{60}},
  \bibinfo{pages}{113006} (\bibinfo{year}{1999}).

\bibitem[{\citenamefont{Maltoni and Stelzer}(2003)}]{Maltoni:2002qb}
\bibinfo{author}{\bibfnamefont{F.}~\bibnamefont{Maltoni}} \bibnamefont{and}
  \bibinfo{author}{\bibfnamefont{T.}~\bibnamefont{Stelzer}},
  \bibinfo{journal}{J. High Energy Phys.} \textbf{\bibinfo{volume}{02}},
  \bibinfo{pages}{027} (\bibinfo{year}{2003}).

\bibitem[{\citenamefont{Abulencia
  \textit{et~al.}}(2006{\natexlab{d}})}]{Abulencia:2005aj}
\bibinfo{author}{\bibfnamefont{A.}~\bibnamefont{Abulencia}}
  \bibnamefont{\textit{et~al.}} (\bibinfo{collaboration}{CDF~Collaboration}),
  \bibinfo{journal}{Phys. Rev. D} \textbf{\bibinfo{volume}{73}},
  \bibinfo{pages}{032003} (\bibinfo{year}{2006}{\natexlab{d}}).

\bibitem[{\citenamefont{Abulencia
  \textit{et~al.}}(2006{\natexlab{e}})}]{Abulencia:2006kv}
\bibinfo{author}{\bibfnamefont{A.}~\bibnamefont{Abulencia}}
  \bibnamefont{\textit{et~al.}} (\bibinfo{collaboration}{CDF~Collaboration})
  (\bibinfo{year}{2006}{\natexlab{e}}), \eprint{hep-ex/0607035}.

\bibitem[{\citenamefont{Cacciari \textit{et~al.}}(2004)\citenamefont{Cacciari,
  Frixione, Mangano, Nason, and Ridolfi}}]{Cacciari:2003fi}
\bibinfo{author}{\bibfnamefont{M.}~\bibnamefont{Cacciari}},
  \bibinfo{author}{\bibfnamefont{S.}~\bibnamefont{Frixione}},
  \bibinfo{author}{\bibfnamefont{M.~L.} \bibnamefont{Mangano}},
  \bibinfo{author}{\bibfnamefont{P.}~\bibnamefont{Nason}}, \bibnamefont{and}
  \bibinfo{author}{\bibfnamefont{G.}~\bibnamefont{Ridolfi}},
  \bibinfo{journal}{J. High Energy Phys.} \textbf{\bibinfo{volume}{04}},
  \bibinfo{pages}{068} (\bibinfo{year}{2004}).

\bibitem[{\citenamefont{Kidonakis and Vogt}(2003)}]{Kidonakis:2003qe}
\bibinfo{author}{\bibfnamefont{N.}~\bibnamefont{Kidonakis}} \bibnamefont{and}
  \bibinfo{author}{\bibfnamefont{R.}~\bibnamefont{Vogt}},
  \bibinfo{journal}{Phys. Rev. D} \textbf{\bibinfo{volume}{68}},
  \bibinfo{pages}{114014} (\bibinfo{year}{2003}).

\bibitem[{\citenamefont{Lai \textit{et~al.}}(2000)}]{Lai:1999wy}
\bibinfo{author}{\bibfnamefont{H.~L.} \bibnamefont{Lai}}
  \bibnamefont{\textit{et~al.}} (\bibinfo{collaboration}{CTEQ Collaboration}),
  \bibinfo{journal}{Eur. Phys. J. C} \textbf{\bibinfo{volume}{12}},
  \bibinfo{pages}{375} (\bibinfo{year}{2000}).

\bibitem[{\citenamefont{Martin \textit{et~al.}}(1998)\citenamefont{Martin,
  Roberts, Stirling, and Thorne}}]{Martin:1998sq}
\bibinfo{author}{\bibfnamefont{A.~D.} \bibnamefont{Martin}},
  \bibinfo{author}{\bibfnamefont{R.~G.} \bibnamefont{Roberts}},
  \bibinfo{author}{\bibfnamefont{W.~J.} \bibnamefont{Stirling}},
  \bibnamefont{and} \bibinfo{author}{\bibfnamefont{R.~S.}
  \bibnamefont{Thorne}}, \bibinfo{journal}{Eur. Phys. J. C}
  \textbf{\bibinfo{volume}{4}}, \bibinfo{pages}{463} (\bibinfo{year}{1998}).

\end{thebibliography}

\end{document}